\documentclass{aa}
\usepackage{graphicx}
\begin{document}

\title{A comparison of the gamma-ray bursts detected by
BATSE and Swift}

\titlerunning{A comparison of the gamma-ray bursts ...}

\author{D. Huja \inst{1}
\and
A. M\'esz\'aros \inst{1}
\and
J. \v{R}\'{\i}pa \inst{1}
 }

\offprints{D. Huja}

\institute{Charles University, Faculty of Mathematics and Physics,
Astronomical Institute,
              V Hole\v{s}ovi\v{c}k\'ach 2, 180 00 Prague 8,
          Czech Republic\\
              \email{David.HUJA@seznam.cz}\\
              \email{meszaros@cesnet.cz}\\
              \email{ripa@sirrah.troja.mff.cuni.cz}
              }

   \date{Received March 18, 2008; accepted  May 22, 2009}

\abstract
{}
{The durations of 388 gamma-ray bursts, detected by the Swift
satellite, are studied statistically in order to search for their 
subgroups. Then the results are compared with
the results obtained earlier from the BATSE database.}
{The standard $\chi^2$ test is used.}
{Similarly to the BATSE database, the short and long subgroups are
well detected also in the Swift data. Also the intermediate subgroup 
is seen in the Swift database.}
{The whole sample of 388 GRBs gives a support for three 
subgroups.}

\keywords{gamma-rays: bursts}

\maketitle

\section{Introduction}

In the years 1991-2000 2704 gamma-ray bursts (GRBs) were
detected by the BATSE instrument onboard the Compton Gamma-Ray
Observatory (\cite{mee01}). After the launch of the
Swift satellite (November 2004) the frequency of detected GRBs  by 
this instrument is cca
100/year (\cite{swift}). Trivially, any comparison of
different databases is highly useful. For example, in the BATSE database -
doubtlessly - three subgroups ("short", "intermediate" and "long" GRBs)
are seen (\cite{ho06,cha07} and references therein). The short
and long subgroups are physically different phenomena (\cite{bal03}).
However, contrary to this, it is still well possible that the intermediate
subgroup is not a real physically different separate subgroup and it is
occurring in the BATSE database due to e.g. some observational biases
arising from the BATSE triggering procedure (\cite{ho06}). The best
choice, to
proceed in this "bias vs. separate subgroup" controversy, is a new
study of another database gained by another instrument. Hence,
it is highly useful to ask: Are these subgroups also seen in the Swift
data-set?

The purpose of this article is the statistical analysis of the Swift
database, which could answer this question.
We will proceed identically to the successful statistical analysis
done on the BATSE Catalog (\cite{ho98}) leading to the discovery of the
third subgroup
(\cite{mu98,bag98,ho99,ha00,rm02,ho02,ho03,bal03,ho06,cha07}).
Recently, a statistical study on the Swift database - using the 
maximum likelihood method - has already shown evidence for the third 
subgroup
(\cite{ho08}). The $\chi^2$ fitting was not used, "because of the small 
population". However, historically, the first evidence for the third 
subgroup in the BATSE database came just from the $\chi^2$ method 
(\cite{ho98}), and also the number of 388 need not be small for this 
testing. Hence, in any case, one has to probe this fitting on the 
Swift data sample too. In addition, since approximately
one third of the Swift's bursts have already well determined redshifts
(contrary to the BATSE's GRBs, where only a few objects had measured
redshifts (\cite{rafe00,nor02,bag03})),
some additional tests can be also done on the samples with and
without redshifts. 

The paper is organized as follows. The samples are defined in Section 2
- these samples are also listed in detail at the end of the article.
Section 3 presents the $\chi^2$ fitting of these samples.
Section 4 discusses the results of this paper and Section 5 summarizes them.

\section{The samples}

We define two samples from the Swift data-set (\cite{swift}): the
sample of GRBs without measured redshifts ($z$)
and the sample with measured redshifts.
These two samples are collected in Tables 4-8 and Tables 9-11,
respectively. We compiled these tables for the convenience;
each table contains the name of GRB,
its BAT duration $T_{90}$, BAT fluence at range
$15-150\; keV$, BAT 1-sec peak photon flux at range $15-150\; 
keV$ and Tables 9-11 also redshift. Only these bursts were taken into
account, of which the GRB duration was measured. The
samples cover the period from November 2004 to the end
of February 2009; the first (last) object is GRB041217 
(GRB090205).
Tables 4-8 (9-11) contain 258 (130) GRBs, and hence the total 
number of GRBs, which are studied in this paper, is 388.

In what follows, we study both samples separately and also
together as one single set ("the whole sample").

\begin{center}
\begin{table} \caption{Results of the $\chi^2$ fitting of the whole sample with 
388 GRBs.}
$$ \begin{array}{lrrrrrrr}
              \hline
Fit & I.   & II.  & III. &  IV. & V. & VI. & VII.   \\
No.\, of &     &      &      &    & & &    \\ 
bins & 30  & 34    & 35     &   36 & 25 & 15 & 31     \\
\hline
1 G &   &     &      &   & & &   \\
\chi_1^2  & 90.5  & 105.6    & 97.4 &  112.3 & 97.5 & 56.6 & 94.5 \\
si.[\%]   & 10^{-6}  & 10^{-8}   &  10^{-6}    &  10^{-8} & 10^{-9}  
& 10^{-5} & 10^{-7}  \\
\mu     & 1.45  &  1.52   & 1.45     &  1.45 & 1.45 & 1.47 & 1.44   \\
\sigma & 0.87     & 0.88  & 0.89    &  0.87 & 0.93 & 0.83 & 0.89 \\
\hline     
2 G &   &     &      &     \\
\chi_2^2  &30.9  & 40.1    &  31.3    &  47.7  & 17.3 & 7.5 & 23.1   \\
si.[\%]   & 15.6  & 6.5    &   34.4   &  2.1  & 
58.8 & 58.6 & 57.1 \\
\mu_1     & 0.48  & 0.46    &  0.41    &  0.32 & 0.06 & 0.42 & 0.33  \\
\sigma_1 & 0.93     & 0.99  &  0.98   &  0.95 & 1.09 & 0.92 & 0.97 \\
\mu_2     & 1.63  &  1.68   &  1.62    &  1.61 & 1.60 & 1.62 & 1.62  \\
\sigma_2 & 0.51     & 0.52  &  0.52   &  0.53 & 0.54 & 0.53 & 0.52 \\
w_2         &0.84     & 0.76   & 0.85    &  0.88 & 0.85 & 0.83 & 0.82   \\
F[\%]         &  10^{-3}    &  10^{-4}  & 10^{-5}  & 10^{-4} & 10^{-5} &  
 10^{-2} & 10^{-6}\\
\hline
3 G &   &     &      &     \\
\chi_3^2  &21.4  &  29.7   &  22.6    &  35.9 & 10.0 & 2.4 & 16.7 \\
si.[\%]   & 43.6  & 23.7   &  65.5 &  11.6  & 
86.7 & 88.2 & 78.0  \\
\mu_1     & 0.34  & 0.91    &  0.11    &  0.19  & -0.01 & 1.07 & 0.28 \\
\sigma_1 & 0.94     & 1.27  &  0.98   &  0.91  & 1.13 & 0.97 & 0.97\\
\mu_2     & 1.19  &  1.18   &  1.24    &   1.04 & 1.06 & 1.12 & 1.62 \\
\sigma_2 & 0.39     & 0.36  &  0.45   & 0.32  & 0.35 & 0.53 & 0.52 \\
w_2         &0.35     & 0.28   & 0.44    &  0.26 &0.29 & 0.30 & 0.18   \\
\mu_3     & 1.91  &  1.93   &  1.94    &  1.84 & 1.85 & 1.84 & 1.84   \\
\sigma_3 & 0.36     & 0.34  & 0.35    &  0.38 & 0.37 & 0.33 & 0.38 \\
w_3         &0.47     & 0.47   & 0.42     &  0.60  & 0.56 & 0.60 & 0.58  
\\
F[\%]         & {\bf 4.04} & {\bf 4.66}  & {\bf 3.07}  & {\bf 4.50} & {\bf 
2.52} & {\bf 3.63} & 5.41  \\
\hline \end{array}$$
\end{table}
\end{center}

\section{$\chi^2$ fitting of the durations}

\subsection{The whole sample}

Since the $\chi^2$ fitting of the GRB duration distribution and the F-test
were successfully used in the work Horv\'ath (1998) 
(presenting the first evidence of the
existence of three GRB subgroups), we proceed identically, 
but with the Swift's data.

The whole sample consists of 388 events  having measured 
$T_{90}$. 
We have fitted the histogram of their decimal $\log T_{90}$ values  
seven  
times (fits I.-VII.). The results are collected in Table 1, 
and the fit No.VI. is seen on Fig. 1. We choose different 
binnings for different fittings with different numbers of bins, with 
different edges of bins, etc. Also the widths of bins are different.
We only require that in each bin the theoretically expected number of GRBs 
should be higher than 5.

At first, the histogram is fitted with one single theo\-retical Gaussian
curve having two free parameters (mean $\mu$ and standard deviation
$\sigma$). The best parameters giving the minimal $\chi^2$ are, e.g., 
for fit No.VI. the following ones:
$\mu = 1.47$, $\sigma = 0.83$ with  $\chi_1^2 = 56.6$. The
goodness-of-fit for 15 - 2 - 1 = 12 degrees of freedom 
(dof) gives the
rejection on the level $10^{-5}\%$ (\cite{TW53,KS73}).
This stands for the rejection of the null-hypothesis 
(i.e. that one Gaussian curve is enough)
that it is correct, because the probability 
of the mistake for this rejection is not higher than $10^{-5}\%$.
The whole sample cannot be described by one single Gaussian curve.
The same is the situation also for the remaining six fittings.

The fitting with the sum of two Gaussian curves (five free parameters:
two means, two standard deviations and one weight $w_2$
(since the first weight is equal to $1-w_2$)) gave for the  
fit No.VI. $\chi^2 = 7.5$. (Note that the value of $w_2$ involves 
that 17\%  (83\%) of GRBs should belong to the short (long) subgroup.)
Here dof = 15 - 5 - 1 = 9 and we
obtained an excellent fit with the significance level 58.6\% (i.e., 
if we suppose that the fit is incorrect, then the probability that  
this assumption is wrong is higher than 58.6\%). The assumption
that the duration distribution is represented by the sum of two 
Gaussian curves cannot be - from the statistical point of view - rejected.
The best fitted curve is also seen on Fig.1, showing a good
correspondence with measured data. Again, the remaining six fits gave 
similar results.

\begin{figure}
\centering
  \includegraphics[width=94mm]{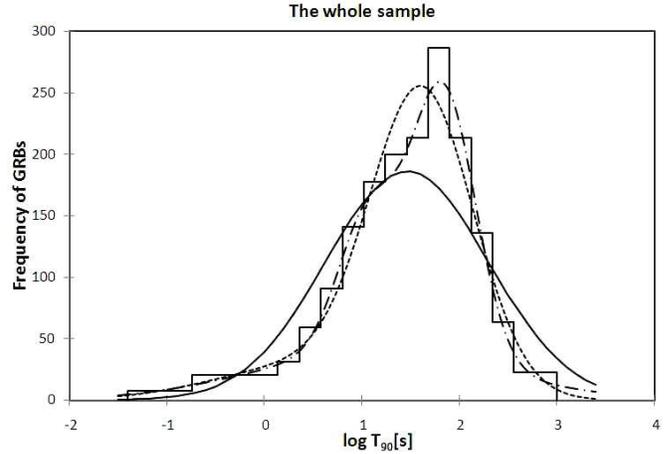}
  \caption{Fitting of the $\log T_{90}$ histogram in the whole sample with 
15 bins (fit No.VI.). The number of GRBs per bin is 
given by the product of the frequency and width. (There are two 
equally populated bins between  $-0.74 < \log 
T_{90} < 0.14$ divided at $\log T_{90} =  -0.30$. For these bins the 
frequencies are 20.45, and hence the number of GRBs in these bins equal to
$20.45\times 0.44 = 9$.) The theoretical curves show the best fits: 
full line = 1 Gaussian curve; dotted line = sum of 2 
Gaussian curves;  dash-dotted line = sum of 3
Gaussian curves.}
  \label{whole}
\end{figure}

   We also performed the fitting with the sum of three Gaussian
curves (eight parameters: three means, three standard deviations and two
independent weights), and obtained an excellent fit with
$\chi_3^2 = 2.4$ for fit No.VI, because the goodness-of-fit gives
for dof = 15 - 8 - 1 = 6 the significance level is 88.2\%.
The best fitted curve is also seen on Fig. 1 showing even better
correspondence with the measured data. The same excellent fits are 
obtained also for the remaining six binnings.

The key question here is following: Is the decreasing
$\Delta \chi^2 =
7.5 - 2.4 = 5.1$ statistically significant? To answer this
question we proceed similarly to Horv\'ath (1998) and used the test
proposed by \cite{ba07} in Appendix A. The significance level from the F-test
is 3.63\%. This implies that the rejection of the null-hypothesis 
(i.e. that the sum of the two Gaussian curves is enough)
is adequate, because the probability of the mistake 
for this rejection is not higher than 3.63\%.
We arrive into a conclusion that the
strengthening of $\chi^2$ need not be a fluctuation. 
Similar results are obtained for the remaining six fits - only for the 
fit No.VII the significance is just above the usual 5\% limit. 
(The significances smaller than 5\% are denoted by boldface.) In other
words, the introduction of the third subgroup - purely from the
statistical point of view - is  significant in six fits from the done 
seven ones. Note that the same F-test can be applied also for the 
difference $\chi_1^2 - \chi_2^2$, and we always obtain the conclusion that
the introduction of the second subgroup - instead of the one single group 
- is strongly supported.

\subsection{The sample with $z$}

\begin{center}
\begin{table} \caption{Results of the $\chi^2$ fitting of the sample with 
the known redshift with  130 GRBs.}
$$ \begin{array}{lrrrrrrr}
              \hline
Fit & I.   & II.  & III. &  IV. & V. & VI. & VII.   \\
No.\, of &     &      &      &    & & &    \\ 
bins & 10  & 11    & 12     &   15 & 16 & 17a & 17b     \\
\hline
1 G &   &     &      &   & & &   \\
\chi_1^2  & 9.7  & 12.2    & 13.9 &  11.8 & 14.8 & 16.8 & 11.9 \\
si.[\%]   & 20.9  & 14.4 & 12.6 &  46.0 &  
31.8 &  26.5 & 61.2  \\
\mu     & 1.55  &  1.52  & 1.53     &  1.69 & 1.54 & 1.54 & 1.53   
\\
\sigma & 0.74     & 0.77  & 0.74    &  0.78 & 0.76 & 0.75 & 0.76 \\
\hline     
2 G &   &     &      &     \\
\chi_2^2  & 2.6  & 6.7    &  7.7   &  4.7  & 7.9 & 10.1 & 5.3   \\
si.[\%]   & 62.4  & 24.0 & 26.0 & 86.0  &  
64.0 & 51.8 & 91.8 \\
\mu_1     & 1.47  & -0.50   &  0.17    &  0.22 & 0.70 & -0.69 & 
-0.66  \\
\sigma_1 & 0.77    & 1.00  &  0.88   &  0.94 & 0.86 & 0.10 & 0.15 \\
\mu_2     & 1.97  &  1.58   &  1.62    &  1.77 & 1.68 & 1.59 & 1.58  
\\
\sigma_2 & 0.10  & 0.63  &  0.58   &  0.60 & 0.55 & 0.62 & 0.63 
\\
w_2         & 0.12 & 0.97   & 0.91    &  0.91 & 0.82 & 0.97 & 
0.97   \\
F[\%]         &  6.97    &  28.21  & 22.19  & {\bf 2.20} & 6.50 &  
 9.74 & {\bf 1.70}\\
\hline
3 G &   &     &      &     \\
\chi_3^2  & 2.0  &  2.0  &  1.9    &  2.2 & 4.0 & 5.4 & 2.8 \\
si.[\%]   & 15.4  & 49.8 & 59.1 & 89.9  &  
77.8 
& 71.4 & 94.8  \\
\mu_1     & 1.04  & -0.13    &  1.16   &  0.17  & 1.07 & -0.57 & 
-0.12 \\
\sigma_1 & 0.30    & 1.00  &  1.05   &  0.50  & 0.92 & 0.05 & 0.14\\
\mu_2     & 1.17  &  0.94  &  1.61    &   1.14 & 1.32 & 1.12 & 0.95 
\\
\sigma_2 & 1.10     & 0.26  &  0.25   & 0.26  & 0.12& 0.42 & 0.33 \\
w_2         & 0.25 & 0.24   & 0.53 &  0.24 &0.49 & 0.48 & 0.28   \\
\mu_3     & 1.94  &  1.85   &  1.91    &  2.00 & 1.88 & 2.00 & 1.86   
\\
\sigma_3 & 0.34     & 0.44  & 0.18    &  0.43 & 0.33 & 0.34 & 0.43 
\\
w_3         & 0.49     & 0.69   & 0.25     &  0.67  & 0.41 & 0.49 & 
0.67 
\\
F[\%]         & 89.64 & 24.41  & 10.51  & 13.29 & 12.66 & 11.40 & 
10.96  
\\
\hline \end{array}$$
\end{table}
\end{center}

The sample contains 130 events with duration informations.
Also here we performed seven fits, but now the number of bins needed 
to be 
smaller due to the smaller number of objects in the sample. Again we 
did different binnings - fits VI. and VII. had 17 bins, but the structure 
was different. In each bin again the number of GRBs was higher than 5. The 
results are collected in Table 2., and fit No.II. is 
shown on Fig. 2.
 
Here the results, compared with the whole sample, are different from 
two reasons. First, here the fittings with one single Gaussian curve are 
also acceptable, and only for two fits the F-test show that the 
introduction of the second subgroup is adequate. Second, the 
introduction of the third subgroup is not needed from the F-test. All 
this shows that this sample can be defined by one single group, and even 
the separation into the short-long GRBs is not needed.

\begin{figure}
\centering
  \includegraphics[width=85mm]{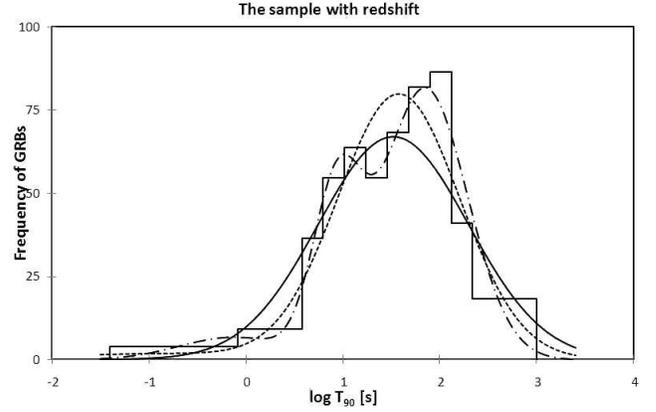}
 \caption{Fitting of $\log T_{90}$ in the sample with known redshifts.
The theoretical curves show the best fits. The notation of the lines
is the same as in Fig.1.
}
  \label{grbplus}
\end{figure}

\subsection{The sample without $z$}

\begin{center}
\begin{table} \caption{Results of the $\chi^2$ fitting of the sample 
without the known redshift with 
258 GRBs.}
$$ \begin{array}{lrrrrrrr}
              \hline
Fit & I.   & II.  & III. &  IV. & V. & VI. & VII.   \\
No.\, of &     &      &      &    & & &    \\ 
bins & 11  & 14    & 16     &   18 & 20 & 22 & 23     \\
\hline
1 G &   &     &      &   & & &   \\
\chi_1^2  & 57.8  & 66.8    & 67.3 &  71.4 & 76.0 & 83.8 & 76.2 \\
si.[\%]   & 10^{-7}  & 10^{-8}    &  10^{-7}    &   
10^{-7} & 10^{-7} & 10^{-8} & 10^{-6}  \\
\mu     & 1.47  &  1.41   & 1.41     &  1.41 & 1.41 & 1.42 & 1.41   \\
\sigma & 0.83     & 0.90  & 0.86    &  0.90 & 0.87 & 0.90 & 0.88 \\
\hline     
2 G &   &     &      &     \\
\chi_2^2  &6.9  & 9.5    &  15.1    &  13.7  & 18.2 & 23.2 & 19.5   \\
si.[\%]   & 22.8  & 30.6 & 13.0 & 31.8  &  
20.0 & 10.9 & 30.0 \\
\mu_1     & -0.04  & 0.09   &  0.36    &  0.22 & 0.38 & 0.28 & 0.48  \\
\sigma_1 & 0.81    & 1.07  &  0.98   &  1.06 & 1.00 & 1.03 & 1.03 \\
\mu_2     & 1.58  &  1.60   &  1.60    &  1.61 & 1.62 & 1.62 & 1.63  \\
\sigma_2 & 0.51     & 0.51  &  0.49   &  0.50 & 0.48 & 0.49 & 0.48 \\
w_2         &0.86     & 0.83   & 0.80    &  0.81 & 0.79 & 0.80 & 0.76   \\
F[\%]         &  0.36    &  0.04  & 0.07  & 0.01 & 0.01 &  
 0.01 & 10^{-3}\\
\hline
3 G &   &     &      &     \\
\chi_3^2  &2.9  &  3.1   &  10.6    &  9.2 & 14.8 & 18.3 & 17.0 \\
si.[\%]   & 23.3  & 68.2 &  15.8 &  42.1  &  19.1 & 14.5 & 25.9  \\
\mu_1     & -0.46  & -0.55    &  -0.52    &  -0.38  & 0.97 & -0.38 & -0.12 
\\
\sigma_1 & 0.40     & 0.82  &  0.59   &  0.80  & 1.13 & 0.78 & 0.86\\
\mu_2     & 0.86  &  1.47   &  1.37    &   1.51 & 1.15 & 1.48 & 1.52 \\
\sigma_2 & 0.33     & 0.56  &  0.57   & 0.57  & 0.19 & 0.56 & 0.55 \\
w_2         &0.20     & 0.77   & 0.67    &  0.78 &0.17 & 0.74 & 0.74   \\
\mu_3     & 1.72  &  1.97   &  1.89    &  1.91 & 1.82 & 1.94 & 1.88   \\
\sigma_3 & 0.43     & 0.15  & 0.28    &  0.15 & 0.29 & 0.20 & 0.20 \\
w_3         &0.72     & 0.13   & 0.25     &  0.11  & 0.45 & 0.15 & 0.13 
\\
F[\%]         & 40.20 & 6.81  & 39.50  & 23.87 & 47.00 & 33.60 & 53.60  \\
\hline \end{array}$$
\end{table}
\end{center}

Here the sample contains 258 events with duration information.
Also here we did seven fits with different
binnings. In each bin again the number of GRBs was higher than 5. The
results are collected in Table 3., and fit No.I. is
shown on Fig. 3.

The results, compared with the whole sample, are similar - except for 
one thing: The
introduction of the third subgroup is not needed from the F-test. All
this shows that this sample can well be defined by the sum of two and only 
two subgroups.

\begin{figure}
\centering
  \includegraphics[width=88mm]{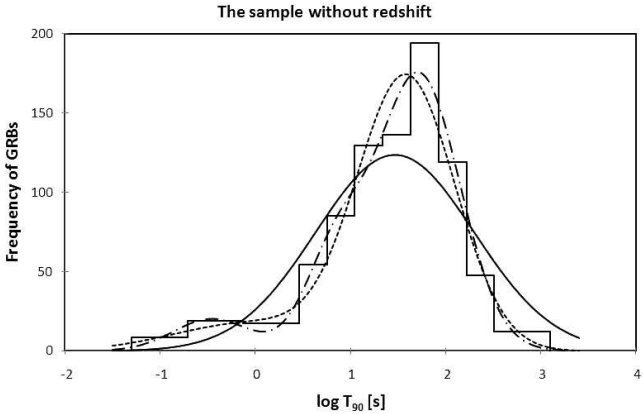}
  \caption{Fitting of $\log T_{90}$ in the sample with unknown
redshifts. The theoretical curves show the best fits.
The notation of the lines
is the same as in Fig.1.
}
  \label{grbminus}
\end{figure}

\section{Discussion of the results}

To discuss the results, first of all, we should recognize
that we have proven the existence of the short and long subgroups
also in the Swift data-set. Both the whole sample and the sample with
no redshifts, respectively, contain these two subgroups, because
the fits with one single Gaussian curve are fully wrong.
It is also highly remarkable that also the weight of the short
subgroup is in accordance with the expectation. As it follows from
Horv\'ath et al. (2006), in the BATSE Catalog the populations of the short,
intermediate and long bursts are roughly in the ratio 20:10:70. Nevertheless,
because the short bursts are harder and Swift is more sensitive to
softer GRBs, one may expect that in the Swift database the population
of short GRBs should be comparable or smaller than 20\% due to 
instrumental 
reasons. The obtained weights for the whole sample (being between 
10 and 26\%) are in accordance with this expectation.
Also the other values of the best parameters - i.e. two
means and two standard deviations - are roughly in the ranges that can be
expected from the BATSE values. The differences can be given by the
different instrumentations. For example, the mean values of the $\log T_{90}$
should be slightly longer in the Swift database compared with the 
BATSE data (\cite{bar05}, \cite{ba06}).
In Horv\'ath (1998) the BATSE's means are -0.35 (short) and 1.52 (long), 
respectively. Here we obtained for the whole sample values from -0.01 to 
0.91 (short) and from 1.60 to 1.94 (long), respectively. All this implies 
that - concerning the short and long GRBs -
the situation is in essence identical to the BATSE data-set.

For the sample
with known redshifts the situation is different, because the fittings
still allow one single Gaussian curve. This result can be easily
explained by selection effects - it is well-known that the observational
determination of the redshifts in the Swift data sample is easier for the 
long bursts  due to observational strategies (simply, it is more 
complicated to detect and to follow the afterglows of 
short GRBs (\cite{swift})).

Concerning the third intermediate subgroup
the whole sample also supports its existence; from 
seven tests six ones gave significances below $5 \%$. 
Hence, strictly speaking, the third subclass does exist and the
probability of the mistake for this claim is not higher than x \%, 
where $2.52 < x < 5.41$. This result is in accordance with the 
expectation, once a comparison with the BATSE database is provided. 
As it was said in Introduction, for the BATSE database the first evidence 
of third subgroup came from this $\chi^2$ method, and hence also for the 
Swift database this test should give positive support for this subgroup, 
if the two datasets are comparable. It is the key result of this article 
that this expectation is fulfilled. Our study has shown that the 
classical $\chi^2$ fitting - in combination with F-test - may
well work also in the Swift database (similarly to the BATSE database
(\cite{ho98})). 

Horv\'ath et al.
(2008) confirmed the third subgroup in the Swift dataset by the 
maximum likelihood (ML) method. Our significance between 2.52\% and
5.41\% is weaker than the 0.46\% significance obtained by Horv\'ath 
et al. (2008), which is expectable, because the ML method is a stronger 
statistical test. This is seen from new two studies, too: the ML test 
on the databases of RHESSI (\cite{ri09}) and BeppoSAX (\cite{ho09}) 
satellites, respectively, 
confirmed the existence of the third intermediate sublass; on the other 
hand, the $\chi^2$ test either did not give a high enough significance
for RHESSI data (\cite{ri09}) or was not used for BeppoSAX data at all 
(\cite{ho09}).

It can also be expected that the mean $\log T_{90}$
for the intermediate group should be much higher in the 
Swift database due to the different redshift distributions
(\cite{ba06,ja06,bag06}). The mean value for the BATSE's 
intermediate subgroup is 0.64 (\cite{ho98}), but here the value is 
between 
1.02 and 1.64. Also Horv\'ath et al. (2008) obtained a similar value
(1.107). Hence, also the typical durations are in accordance with the 
expectations.  

The sample with no redshift did not find the third subgroup. This result 
can be explained by the smaller number of objects in the sample. The 
sample with known redshifts is strongly biased by selection effects, and 
here even the existence of the short subgroup was in doubt - hence, it 
seems to be hopeless to obtain some conclusions concerning the third 
subgroup.

\section{Conclusions}

Since the $\chi^2$ fitting of the GRB duration distribution and the F-test
were successfully used in the work \cite{ho98} (presenting the 
first evidence of the
existence of three GRB subgroups), we proceed identically, 
but with the Swift's data.

The results may be summarized in the following four points:

1. Concerning the short and long subgroups all is
in accordance with the expectation: they are detected also in the Swift
database and - in addition - in the Swift database the weight of the
short subgroup is smaller, which can be well explained by the Swift's
higher effective sensitivity to the softer bursts.

2. The whole sample of 388 objects gives support for three 
subgroups,
because from seven fittings of the whole sample six ones  
confirmed the existence of the intermediate subgroup on a smaller than 5\%
significance level. Hence, concerning the Swift database, the 
situation is similar to the BATSE dataset  - although our signficances 
are weaker than $> 0.02\%$ of Horv\'ath (1998).

3. The samples with and without known redshifts separately are either not 
enough populated, or strongly biased. Hence, no far reaching 
conclusions can be drawn from them. 

4. Similarly to the BATSE database, here it is shown again that the
classical $\chi^2$ test - in
combination with F-test - is also effective for the Swift GRB sample.

\begin{acknowledgements}

Thanks are due to valuable discussions with
Z. Bagoly, L.G. Bal\'azs, I. Horv\'ath and P. Veres.
This study was supported by the GAUK grant No. 46307, by the OTKA
grants No. T48870 and K77795,
by the Grant Agency of the Czech Republic grant No. 205/08/H005,
and by the Research Program MSM0021620860 of the Ministry
of Education of the Czech Republic. 
The useful remarks of the referee, C. Guidorzi, are kindly 
acknowledged.

\end{acknowledgements}

\begin{center}
\begin{table} \caption{Swift GRBs with no measured redshifts;
Part I.}
$$ \begin{array}{lrrr}
              \hline
GRB & T_{90}   & fluence  & peak-flux   \\
    &  sec   & 10^{-7}\,erg/cm^2   & ph/(cm^2 sec) \\
\hline
090201  &        83.0        &        300.00        &        14.70\\
090129        &        17.5        &        21.00        &        3.70\\
090123        &        131.0        &        29.00        &        1.70\\
090118        &        16.0        &        4.00        &        n/a\\
090113        &        9.1        &        7.60        &        2.50\\
090111        &        24.8        &        6.20        &        0.90\\
090107A        &        12.2        &        2.30        &        1.10\\
081230        &        60.7        &        8.20        &        0.70\\
081228        &        3.0        &        0.89        &        0.60\\
081226A        &        0.4        &        0.99        &        2.40\\
081221        &        34.0        &        181.00        &        18.20\\
081211A        &        3.5        &        1.30        &        0.80\\
081210        &        146.0        &        18.00        &        2.50\\
081203B        &        23.4        &        21.00        &        n/a\\
081128        &        100.0        &        23.00        &        1.30\\
081127        &        37.0        &        4.90        &        0.60\\
081126        &        54.0        &        33.00        &        3.70\\
081109A        &        190.0        &        36.00        &        1.10\\
081104        &        59.1        &        20.00        &        1.00\\
081102        &        63.0        &        23.00        &        1.40\\
081101        &        0.2        &        0.62        &        3.60\\
081025        &        23.0        &        19.00        &        1.30\\
081024A        &        1.8        &        1.20        &        1.10\\
081022        &        160.0        &        25.00        &        0.60\\
081017        &        320.0        &        14.00        &        0.07\\
081016B        &        2.6        &        0.99        &        0.50\\
081012        &        29.0        &        11.00        &        1.00\\
081011        &        9.0        &        1.60        &        0.40\\
080919        &        0.6        &        0.72        &        1.20\\
080916B        &        32.0        &        6.30        &        0.60\\
080915B        &        3.9        &        9.90        &        8.50\\
080915A        &        14.0        &        2.30        &        0.50\\
080905A        &        1.0        &        1.40        &        1.30\\
080903        &        66.0        &        14.00        &        0.80\\
080822B        &        64.0        &        1.70        &        0.06\\
080802        &        176.0        &        13.00        &        0.30\\
080727C        &        79.7        &        52.00        &        2.30\\
080727B        &        8.6        &        31.00        &        7.60\\
080727A        &        4.9        &        1.30        &        0.30\\
080725        &        120.0        &        37.00        &        2.30\\
080723A        &        17.3        &        3.30        &        0.90\\
080714        &        33.0        &        25.00        &        4.20\\
080703        &        3.4        &        2.00        &        1.00\\
080702B        &        20.0        &        5.00        &        0.50\\
080702A        &        0.5        &        0.36        &        0.70\\
080701        &        18.0        &        7.10        &        2.20\\
\hline \end{array}$$
\end{table}
\end{center}

\begin{center}
\begin{table} \caption{Swift GRBs with no measured redshifts;
Part II.}
$$ \begin{array}{lrrr}
              \hline
GRB & T_{90}   & fluence  & peak-flux   \\
    &  sec   & 10^{-7}\,erg/cm^2   & ph/(cm^2 sec) \\
\hline
080623        &        15.2        &        10.00        &        2.00\\
080613B        &        105.0        &        58.00        &        2.70\\
080602        &        74.0        &        32.00        &        2.90\\
080524        &        9.0        &        2.90        &        0.40\\
080523        &        102.0        &        8.80        &        0.50\\
080517        &        64.6        &        5.60        &        0.60\\
080515        &        21.0        &        20.00        &        3.90\\
080506        &        150.0        &        13.00        &        0.40\\
080503        &        170.0        &        20.00        &        0.90\\
080426        &        1.7        &        3.70        &        4.80\\
080409        &        20.2        &        6.10        &        3.70\\
080405        &        40.0        &        12.00        &        n/a\\
080328        &        90.6        &        94.00        &        5.50\\
080325        &        128.4        &        49.00        &        1.40\\
080320        &        14.0        &        2.70        &        0.60\\
080319D        &        24.0        &        3.20        &        0.10\\
080319A        &        64.0        &        48.00        &        1.20\\
080315        &        65.0        &        1.40        &        0.04\\
080307        &        125.9        &        8.70        &        0.40\\
080303        &        67.0        &        6.60        &        1.40\\
080229A        &        64.0        &        90.00        &        5.70\\
080218B        &        6.2        &        5.10        &        3.10\\
080218A        &        27.6        &        6.30        &        1.40\\
080212        &        123.0        &        29.00        &        1.20\\
080207        &        340.0        &        61.00        &        1.00\\
080205        &        106.5        &        21.00        &        1.40\\
080130        &        65.0        &        7.70        &        0.20\\
080129        &        48.0        &        8.90        &        0.20\\
080123        &        115.0        &        5.70        &        1.80\\
080121        &        0.7        &        0.30        &        n/a\\
071129  & 420.0   & 35.00 & 0.90 \\
071118  & 71.0 & 5.00 & 0.30 \\
071112B & 0.3 & 0.48 & 1.30 \\
07110 & 19.0 &  0.76 & 0.40 \\
071028B & 55.0 & 2.50 & 1.40 \\
071028A & 27.0 & 3.00 & 0.30 \\
071025   & 109.0 & 65.00 & 1.60 \\
071018   &376.0 & 10.00 & 0.20  \\
071013   & 26.0 & 3.20 & 0.40 \\
071011   & 61.0 & 22.00 & 1.70 \\
071008 & 18.0 & 2.40 & 0.50 \\
071006   & 50.0 & 1.40 & 13.00 \\
071001 & 58.5 & 7.70 & 0.90 \\
070923   & 0.1 & 0.10 & 2.40 \\
070920B & 20.2 & 6.60 & 0.80 \\
070920A & 56.0 & 5.10 & 0.30 \\
070917   & 7.3 & 20.00 & 8.50 \\
070913   & 3.2 & 2.50 & 1.40 \\
070911 & 162.0 & 120.00 & 3.90 \\
\hline \end{array}$$
\end{table}
\end{center}

\begin{center}
\begin{table} \caption{Swift GRBs with no measured redshifts;
Part III.}
$$ \begin{array}{lrrr}
              \hline
GRB & T_{90}   & fluence  & peak-flux   \\
    &  sec   & 10^{-7}\,erg/cm^2   & ph/(cm^2 sec) \\
\hline
070810B & 80.0 & 0.12 & 1.80 \\
070809 & 1.3 &  1.00 & 1.20 \\
070808 & 32.0 & 12.00 & 2.00 \\
070805 & 31.0 & 7.20 & 0.70 \\
070731 & 2.9 &  1.60 & 1.20 \\
070729 & 0.9 & 1.00 & 1.00 \\
070721A & 3.4 & 0.71 & 0.70 \\
070714A & 2.0 & 1.50 & 1.80 \\
070704   & 380.0 & 59.00 & 2.10 \\
070628   & 39.1 & 35.00 & 5.10 \\
070621   & 33.3 & 43.00 & 2.50 \\
070616 & 402.0 & 192.00 & 1.90 \\
070612B & 13.5 & 17.00 & 2.60 \\
070610 & 4.6 &  2.40 & 0.90 \\
070531 & 44.0 & 11.00 & 1.00 \\
070520B & 66.0 & 9.20 & 0.40 \\
070520A & 18.0 & 2.50 & 0.40 \\
070518 & 5.5 & 1.60 & 0.70 \\
070517 & 9.0 &  2.60 & 0.80 \\
070509 & 7.7 & 1.70 &   0.70 \\
070429B & 0.5 & 0.63 & 1.80 \\
070429A & 163.0 & 9.20 & 0.40 \\
070427   & 11.0 & 7.20 & 1.30 \\
070420   & 77.0 & 140.00 & 7.10 \\
070419B & 236.5 & 75.00 & 1.40 \\
070412 & 34.0 & 4.80 & 0.70 \\
070406 & 0.7 & 0.45 & 0.70 \\
070330 & 9.0 & 1.80 & 0.90 \\
070328 & 69.0 & 89.00 & 4.20 \\
070227 & 7.0 &  16.00 & 2.70 \\
070224 & 34.0 & 3.10 & 0.30 \\
070223 & 89.0 & 17.00 & 0.70 \\
070220   & 129.0 & 106.00 & 5.88 \\
070219 & 17.0 & 3.20 & 0.70 \\
070209 & 0.1 & 0.11 & 2.40 \\
070129 & 460.0 & 31.00 & 0.60 \\
070126 & 51.0 & 1.60 & 0.20 \\
070103 & 19.0 & 3.40 & 1.10 \\
061222A & 72.0 & 83.00 & 9.20 \\
061218 & 4.1 & 0.20 & 0.41 \\
061202   & 91.0 & 35.00 & 2.60 \\
061126 & 191.0 & 72.00 & 9.80 \\
061102 & 17.6 & 1.90 & 0.20 \\
061028 & 106.0 & 9.70 & 0.70 \\
061027 & 150.0 & 4.70 & 0.08 \\
061021 & 46.0 & 30.00 & 6.10 \\
061019 & 191.0 & 17.00 & 2.20 \\
061006 & 130.0 & 14.30 & 5.36 \\
061004 & 6.2 & 5.70 & 2.50 \\
061002   & 17.6 & 6.80 & 0.80 \\
060929 & 12.4 & 2.80 & 0.40 \\
060923C & 76.0 & 16.00 & 1.00 \\
060923B & 8.8 & 4.80 & 1.50 \\
060923A & 51.7 & 8.70 & 1.30 \\
060919 & 9.1 & 5.50 & 2.20 \\
060904A & 80.0 & 79.00 & 4.90 \\
\hline \end{array}$$
\end{table}
\end{center}

\begin{center}
\begin{table} \caption{Swift GRBs with no measured redshifts;
Part IV.}
$$ \begin{array}{lrrr}
              \hline
GRB & T_{90}   & fluence  & peak-flux   \\
    &  sec   & 10^{-7}\,erg/cm^2   & ph/(cm^2 sec) \\
 \hline
060825   & 8.1 & 9.80 & 2.70 \\
060813 & 14.9 & 55.00 & 9.00 \\
060807 & 34.0 & 7.30 & 0.80 \\
060805 & 5.4 & 0.74 & 0.30 \\
060804 & 16.0 & 5.10 & 1.20 \\
060801 & 0.5 & 0.81 & 1.30 \\
060728 & 60.0 & 2.40 & n/a \\
060719 & 55.0 & 16.00 & 2.30 \\
060717 & 3.0 & 0.65 & 0.50 \\
060712   & 26.0 & 13.00 & 1.70 \\
060607B & 31.0 & 17.00 & 1.50 \\
060602A & 60.0 & 16.00 & 0.50 \\
060516   & 160.0 & 11.00 & 0.30 \\
060515   & 52.0 & 14.00 & 0.80 \\
060510A & 21.0 & 98.00 & 17.00 \\
060507   & 185.0 & 45.00 & 1.30 \\
060501 & 26.0 & 12.00 & 1.90 \\
060428B & 58.0 & 7.20 & 0.60 \\
060428A & 39.4 & 14.00 & 2.40 \\
060427   & 64.0 & 5.00 & 0.30 \\
060424   & 37.0 & 6.80 & 1.60 \\
060421   & 11.0 & 12.00 & 3.00 \\
060413  1 & 50.0 & 36.00 & 0.90 \\
060403 & 30.0 & 14.00 & 1.00 \\
060323   & 18.0 & 5.70 & 0.80 \\
060322 & 213.0 & 51.00 & 2.10 \\
060319   & 12.0 & 2.70 & 1.10 \\
060313 & 0.7 & 11.30 & 12.10 \\
060312   & 43.0 & 18.00 & 1.50 \\
060306   & 61.0 & 22.00 & 6.10 \\
060223B & 10.2 & 16.00 & 2.90 \\
060219   & 62.0 & 4.20 & 0.60 \\
060211B & 29.0 & 4.70 & 0.70 \\
060211A & 126.0 & 15.00 & 0.40 \\
060204B & 134.0 & 30.00 & 1.30 \\
060203   & 60.0 & 8.50 & 0.60 \\
060202   & 203.7 & 24.00 & 0.50 \\
060117   & 16.0 & 204.00 & 48.90 \\
060111B & 59.0 & 16.00 & 1.40 \\
060111A & 13.0 & 11.80 & 1.72 \\
060110   & 17.0 & 14.00 & 1.90 \\
060109   & 116.0 & 6.40 & 0.50 \\
060105   & 55.0 & 182.00 & 7.50 \\
060102 & 21.0 & 2.40 & 0.40 \\
051227 & 8.0 & 2.30 & 0.97 \\
051221B & 61.0 & 11.30 & 0.54 \\
051213   & 70.0 & 8.00 & 0.51 \\
051210 & 1.4 & 0.83 & 0.75 \\
051117B & 8.0 & 1.40 & 0.46 \\
051117A & 140.0 & 46.00 & 0.93 \\
051114   & 2.2 & 1.32 & 0.73 \\
051113 & 94.0 & 26.00 & 2.40 \\
051105A & 0.3 & 0.20 & 2.00 \\
051021B & 47.0 & 9.10 & 0.63 \\
051016A & 22.0 & 8.80 & 1.60 \\
051012   & 13.0 & 2.90 & 0.60 \\
051008   & 16.0 & 58.00 & 5.50 \\
051006   & 26.0 & 12.80 & 1.90 \\
051001   & 190.0 & 18.00 & 0.51 \\
\hline \end{array}$$
\end{table}
\end{center}

\begin{center}
\begin{table} \caption{Swift GRBs with no measured redshifts;
Part V.}
$$ \begin{array}{lrrr}
              \hline
GRB & T_{90}   & fluence  & peak-flux   \\
    &  sec   & 10^{-7}\,erg/cm^2   & ph/(cm^2 sec) \\
 \hline
050925 & 0.1 & 0.75 & 1.50 \\
050922B & 250.0 & 26.00 & 1.02 \\
050916   & 90.0 & 11.00 & 0.69 \\
050915B & 40.0 & 34.00 & 2.34 \\
050915A & 53.0 & 8.80 & 0.80 \\
050911 & 16.0 & 3.01 & 1.31 \\
050906 & 0.1 &  0.07 & 1.15 \\
050827   & 49.0 & 21.20 & 1.86 \\
050822   & 102.0 & 26.10 & 2.47 \\
050820B & 13.0 & 21.20 & 4.06 \\
050819   & 36.0 & 3.52 & 0.39 \\
050815   & 2.8 & 0.92 & 0.56 \\
050801   & 20.0 & 3.12 & 1.47 \\
050721   & 39.0 & 30.80 & 3.08 \\
050717   & 86.0 & 61.70 & 6.34 \\
050715   & 52.0 & 14.40 & 1.07 \\
050713A & 70.0 & 52.50 & 4.78 \\
050712 & 48.0 & 11.00 & 0.55 \\
050701   & 22.0 & 13.60 & 2.77 \\
050607   & 26.5 & 6.05 & 0.99 \\
050528   & 10.8 & 4.40 & 1.23 \\
050509A & 11.6 & 3.38 & 0.88 \\
050502B & 17.5 & 4.72 & 1.43 \\
050422   & 59.2 & 6.15 & 0.57 \\
050421   & 10.3 & 1.18 & 0.44 \\
050418 & 83.0 & 53.90 & 3.80 \\
050416B & 5.4 & 11.30 & 5.85 \\
050412 & 26.0 & 5.66 & 0.49 \\
050410 & 43.0 & 43.00 & 1.80 \\
050326   & 29.5 & 90.50  & 12.40 \\
050306   & 160.0 & 120.00 & 3.64 \\
050219B & 27.0 & 164.00 & 25.40 \\
050219A & 23.0 & 42.10 & 3.61 \\
050215B & 8.0 & 2.33 & 0.68 \\
050215A & 6.0 & 7.29 & 0.51 \\
050202 & 0.1 & 0.33 & 2.98 \\
050128   & 13.8 & 51.70 & 7.59 \\
050124   & 4.1 & 12.30 & 5.57 \\
050117  & 169.0 & 91.20 & 2.40 \\
041228 & 62.0 & 36.10 & 1.65 \\
041226 & 15.0 & 3.23 & 0.34 \\
041224  & 235.0 & 75.30 & 2.95 \\
041223  & 107.0 & 171.00 & 7.49 \\
041220 & 5.0 & 3.82 & 1.83 \\
041219C & 40.0 & 20.00 & 1.50 \\
041219B & 30.0 & n/a & 10.00 \\
041219A & 520.0 & 1000.00 & 25.00 \\
041217  & 7.5 &  65.70 & 4.40 \\
\hline \end{array}$$
 \end{table}
\end{center}

\begin{center}
\begin{table} \caption{Swift GRBs with known redshifts; Part I.}
$$ \begin{array}{lrrrr}
              \hline
GRB & T_{90}   & fluence  & peak-flux & z  \\
    &  sec   & 10^{-7}\,erg/cm^2   & ph/(cm^2 sec) & \\
\hline
090205    &        8.8        &        1.90  &     0.50   &  4.6749   \\
090102        &        27.0        &        0.68   &  5.50   &   1.5477   \\
081222        &        24.0        &        48.00   &   7.70   &  2.7467 \\
081203A &   294.0        &        77.00        &        2.90 &   2.1000  \\
081121    &  14.0        &        41.00        &        4.40  &  2.5120   \\
081118     &        67.0        &        12.00        &  0.60  & 2.5800   \\
081029    &      270.0        &        21.00        &     0.50  &   3.8474  \\
081028A    &        260.0   &        37.00        &   0.50  &    3.03800 \\
081008    &    185.5   &        43.00        &        1.30  &   1.96775  \\
081007    &        10.0        &        7.10   &  2.60  &    0.52950    \\
080928    &   280.0        &        25.00        &        2.10  &  1.6910 \\
080916A   &    60.0        &        40.00   &  2.70  &        0.68900    \\
080913    &    8.0        &        5.60        &        1.40   &  6.57000 \\
080906    &    147.0        &        35.00        &   1.00   &   2.00000  \\
080905B   &        128.0        &   18.00 &   0.50    &  2.37400   \\
080810    &        106.0        &        46.00        &   2.00 &  3.35000  \\
080805    &     78.0        &        25.00        &  1.10  &   1.50500  \\
080804    &        34.0        & 36.00  &  3.10  & 2.20225        \\
080721    &    16.2        &        120.00        &  20.90  &  2.59650 \\
080710    &    120.0   &   14.00   &  1.00  &        0.84500        \\
080707    &        27.1    &        5.20        &   1.00  &    1.23000  \\
080607    &   79.0        &        240.00        &   23.10   &   3.03600  \\
080605    &    20.0    &   133.00        &        19.90     &  1.63980  \\
080604    &    82.0    &    8.00   &        0.40   &   1.41600   \\
080603B   &    60.0  &   24.00   &   3.50   &    2.69000        \\
080520    &    2.8   &   0.55   &   0.50   &    1.54500     \\
080516    &    5.8   &   2.60   &   1.80   &     3.20000        \\
080430    &   16.2   &        12.00        &        2.60   &   0.75850 \\
080413B   &   8.0    &     32.00        &        18.70  &   1.10000    \\
080413A   &        46.0   &   35.00     &   5.60   &   2.43300   \\
080411    &   56.0        &        264.00        &        43.20 &  1.03000  \\
080330    &   61.0   &        3.40        &        0.90   &  1.51000  \\
080319C   &   34.0   &  36.00  &        5.20  &        1.95000    \\
080319B   &   50.0   &        810.00        &    24.80  &    0.93700   \\
080310    &   365.0        &        23.00  &   1.30 &        2.42580    \\
080210    &   45.0   &  18.00   &    1.60  &   2.64100    \\
\hline \end{array}$$
\end{table}
\end{center}

\begin{center}
\begin{table} \caption{Swift GRBs with known redshifts; Part II.}
$$ \begin{array}{lrrrr}
              \hline
GRB & T_{90}   & fluence  & peak-flux & z  \\
    &  sec   & 10^{-7}\,erg/cm^2   & ph/(cm^2 sec) & \\
\hline
071227 & 1.8 & 2.20 & 1.60 & 0.3835 \\
071122  & 68.7 & 5.80 & 0.40 & 1.14 \\
071117  & 6.6 & 24.00 & 11.30 & 1.331 \\
071112C & 15.0 & 30.00 & 8.00 & 0.82 \\
071031 & 180.0 & 9.00 & 0.50 & 2.692 \\
071021 & 225.0 & 13.00 & 0.70 & 5.0  \\
071020 & 4.2 &  23.00 &  8.40 & 2.1435 \\
071010B & 35.7 & 44.00 & 7.70 & 0.947 \\
071010A & 6.0 & 2.00 & 0.80 &   0.98 \\
071003   & 150.0 & 83.00 & 6.30 & 1.0185 \\
070810A  & 11.0 & 6.90 & 1.90 & 2.17 \\
070802 & 16.4 & 2.80 & 0.40 & 2.45 \\
070724A & 0.4 & 0.80 & 1.00 & 0.457 \\
070721B & 340.0 & 36.00 & 1.50 & 3.626 \\
070714B & 64.0 & 7.20 & 2.70 & 0.92 \\
070612A & 370.0 & 110.00 & 1.50 & 0.617 \\
070611 & 12.0 & 39.00 & 0.80 & 2.04 \\
070529 & 109.0 & 26.00 & 1.40 & 2.4996 \\
070521 & 37.9 & 80.00 & 6.70 & 0.553 \\
070508   & 21.0 & 200.00 & 24.70 & 0.82 \\
070506   & 4.3 & 2.10 & 1.00 & 2.31 \\
070419A & 116.0 & 5.60 & 2.80 & 0.97 \\
070411 & 101.0 & 25.00 & 1.00 & 2.954 \\
070318 & 63.0 & 23.00 & 1.60 & 0.838 \\
070306 & 210.0 & 55.00 & 4.20 & 1.497 \\
070208 & 48.0 & 4.30 & 0.90 & 1.165 \\
070110   & 85.0 & 16.00 & 0.60 & 2.352 \\
061222B & 40.0 & 22.00 & 1.50 & 3.355 \\
061217 & 0.3 & 0.46 & 1.30 & 0.827 \\
061210   & 85.0 & 11.00 & 5.30 & 0.41 \\
061201 & 0.8 & 3.30 & 3.90 & 0.111 \\
061121 & 81.0 & 137.00 & 21.10 & 1.314 \\
061110B & 128.0 & 13.00 & 0.40 & 3.44 \\
061110A & 41.0 & 11.00 & 0.50 & 0.758 \\
 061007 & 75.0 & 450.00 & 15.30 & 1.2615 \\
060927 & 22.6 & 11.00 & 2.80 & 5.6 \\
060926 & 8.0 & 2.20 & 1.10 & 3.208 \\
060912 & 5.0 & 13.00 & 8.50 & 0.937 \\
060908 & 19.3 & 29.00 & 3.20 & 2.43 \\
060906 & 43.6 & 22.10 & 2.00 & 3.685 \\
060904B & 192.0 & 17.00 & 2.50 & 0.703 \\
060814 & 146.0 & 150.00 & 7.40 & 0.84 \\
060729   & 116.0 & 27.00 & 1.40 & 0.54 \\
060714   & 115.0 & 30.00 & 1.40 & 2.71 \\
060708 & 9.8 & 5.00 & 2.00 & 2.3 \\
060707   & 68.0 & 17.00 & 1.10 & 3.43 \\
\hline \end{array}$$
\end{table}
\end{center}

\begin{center}
\begin{table} \caption{Swift GRBs with known redshifts; Part III.}
$$ \begin{array}{lrrrr}
              \hline
GRB & T_{90}   & fluence  & peak-flux & z  \\
    &  sec   & 10^{-7}\,erg/cm^2   & ph/(cm^2 sec) & \\
\hline
060614   & 102.0 & 217.00 & 11.60 & 0.1275 \\
060607A & 100.0 & 26.00 & 1.40 & 3.082 \\
060605   & 15.0 & 4.60 & 0.50 & 3.76 \\
060604   & 10.0 & 1.30 & 0.60 & 2.68 \\
060526 & 13.8 & 4.90 & 1.70 & 3.21 \\
060522 & 69.0 & 11.00 & 0.60 & 5.11 \\
060512   & 8.6 & 2.30 & 0.90 & 0.4428 \\
060510B & 276.0 & 42.00 & 0.60 & 4.9 \\
060505   & 4.0 & 6.20 & 1.90 & 0.089 \\
060502B & 90.0 & 0.40 & 4.40 & 0.287 \\
060502A & 33.0 & 22.00 & 1.70 & 1.51 \\
060418   & 52.0 & 81.00 & 6.70 & 1.4895 \\
060223A & 11.0 & 6.80 & 1.40 & 4.41 \\
060210   & 255.0 & 77.00 & 2.80 & 3.91 \\
060206   & 7.0 & 8.40 & 2.80 & 4.048 \\
060123   & 900.0 & 3.00 & 0.04 & 1.099 \\
060116   & 113.0 & 26.00 & 1.10 & 5.3 \\
060115   & 142.0 & 18.00 & 0.90 & 3.53 \\
060108   & 14.4 & 3.70 & 0.70 & 2.03 \\
051221A & 1.4 & 11.60 & 12.10 & 0.547 \\
051111 & 47.0 & 39.00 & 2.50 & 1.549 \\
051109B & 15.0 & 2.70 & 0.50 & 0.08 \\
051109A & 36.0 & 21.00 & 3.70 & 2.346 \\
051016B & 4.0 & 1.70 & 1.32 & 0.9364 \\
050922C & 5.0 & 17.00 & 7.36 & 2.198 \\
050908   & 20.0 & 4.91 & 0.70 & 3.3459 \\
050904   & 225.0 & 50.70 & 0.66 & 6.29 \\
050826   & 35.0 & 4.51 & 0.43 &  0.297 \\
050824 & 25.0 & 2.92 & 0.52 & 0.83 \\
050820A & 26.0 & 40.10 & 2.50 & 2.61335 \\
050814 & 65.0 & 18.30 & 0.75 & 5.3 \\
050813   & 0.6 & 0.43 & 0.92 & 1.80 \\
050803 & 85.0 & 22.30 & 1.08 & 0.422 \\
050802   & 13.0 & 22.00 & 2.65 & 1.71 \\
050730   & 155.0 & 24.20 & 0.57 & 3.9669 \\
050724 & 3.0 &  11.80 & 3.29 & 0.2575 \\
050603   & 13.0 & 76.30 & 27.60 & 2.821 \\
050525A & 8.8 & 156.00 & 42.30 & 0.606 \\
050509B  & 0.04 & 0.13 & 1.33 & 0.225 \\
050505 & 60.0 & 24.90 & 1.81 & 4.27 \\
050416A & 2.4 & 4.31 & 4.97 & 0.6535 \\
050406 & 3.0 & 0.81 & 0.38 & 2.44 \\
050401 & 33.0 & 85.50 & 12.60 & 2.9 \\
050319   & 10.0 & 6.25 & 1.45 & 3.24 \\
050318   & 32.0 & 13.10 & 3.20 & 1.44 \\
050315   & 96.0 & 32.30 & 1.98 & 1.949 \\
050223   & 23.0 & 6.40 & 0.70 & 0.58775 \\
050126   & 26.0 & 8.60 & 0.70 & 1.29 \\
\hline \end{array}$$
 \end{table}
\end{center}


\begin{thebibliography}{}

\bibitem[Bagoly et al. 1998]{bag98} Bagoly, Z., M\'esz\'aros, A.,
Horv\'ath, I., Bal\'azs, L.G. \& M\'esz\'aros, P. 1998, ApJ, 498, 342

\bibitem[Bagoly et al. 2003]{bag03} Bagoly, Z., Csabai, I.,
M\'esz\'aros, A., M\'esz\'aros, P., Horv\'ath, I.,
Bal\'azs, L.G. \& Vavrek I. 2003,  A\&A, 398, 919

\bibitem[Bagoly et al. 2006]{bag06} Bagoly, Z., M\'esz\'aros, A., 
Bal\'azs, L.G., Horv\'ath, I., Klose, S., 
Larsson, S., M\'esz\'aros, P., Ryde, F., Tusn\'ady 2006,  A\&A, 453, 797

\bibitem[Bal\'azs et al. 2003]{bal03}
Bal\'azs, L.G., Bagoly, Z., Horv\'ath, I., M\'esz\'aros, A.  \&
M\'esz\'aros, P.  2003, A\&A, 401, 129

\bibitem[Band et al. (1997)]{ba07} Band, D.L., Ford, L.A., Matteson,
J.L, Briggs, M.S., Paciesas, W.S., Pendleton, G.N. \& Preece, R.D. 1997,
ApJ, 485, 747

\bibitem[Band 2006]{ba06} Band, D.L. 2006, ApJ, 644, 378

\bibitem[Barthelmy et al. 2005]{bar05} 
Barthelmy, S.D., et al. 2005, Nature, 438, 994

\bibitem[Chattopadhyay et al. 2007]{cha07} Chattopadhyay, T., Misra, R.,
Chattopadhyay, A.K.\& Naskar, M. 2007, ApJ, 667, 1017

\bibitem[Gehrels et al. 2005]{swift} Gehrels, N., et al. (Swift team)
2005, 
http://heasarc.gsfc.nasa.gov/docs/swift/archive/grb$\underline{\;\;}$table/

\bibitem[Hakkila et al. 2000]{ha00} Hakkila, J. et al. 2000, ApJ, 538,
165

\bibitem[Horv\'ath 1998]{ho98} Horv\'ath, I. 1998, ApJ, 508, 757

\bibitem[Horv\'ath 1999]{ho99} Horv\'ath, I. 1999,
J. Korean Astron. Soc., 35, S629

\bibitem[Horv\'ath 2002]{ho02} Horv\'ath, I. 2002, A\&A, 392, 791

\bibitem[Horv\'ath 2003]{ho03} Horv\'ath, I. 2003, in Statistical
Challenges in Modern Astronomy III.,  ed.\ E. D. Feigelson, \& G. Jogesh Babu,
(Springer, Berlin) 439

\bibitem[Horv\'ath et al. 2006]{ho06} Horv\'ath, I., Bal\'azs, L.G.,
Bagoly, Z., Ryde, F. \&  M\'esz\'aros, A.  2006, A\&A, 447, 23

\bibitem[Horv\'ath et al. 2008]{ho08} Horv\'ath, I., Bal\'azs, L.G.,
Bagoly, Z. \& Veres, P.  2008, A\&A, 489, L1

\bibitem[Horv\'ath 2009]{ho09} Horv\'ath, I.  2009, ASS, in press; 
astro-ph/0905.0860

\bibitem[Jakobsson et al. 2006]{ja06}Jakobsson, P., et al. 2006, A\&A, 447, 897

\bibitem[Kendall \& Stuart 1973]{KS73} Kendall, M.G., \&
Stuart, A.  1973, The Advanced Theory of Statistics, Charles
Griffin \& Co. Ltd., London \& High Wycombe

\bibitem[Meegan et al. 2001]{mee01}
Meegan, C.A., et al. 2001, Current BATSE Gamma-Ray Burst Catalog,
http://gammaray.msfc.nasa.gov/batse/grb/catalog

\bibitem[M\'esz\'aros et al. (2000)]{me00} M\'esz\'aros, A.,
Bagoly, Z., Horv\'ath, I.,
Bal\'azs, L.G. \& Vavrek, R. 2000, ApJ, 539, 98

\bibitem[Mukherjee et al. 1998]{mu98} Mukherjee, S., Feigelson, E. D.,
Babu, G. J., Murtagh, F., Fraley, C. \& Raftery, A. 1998, ApJ, 508, 314

\bibitem[Norris 2002]{nor02} Norris, J.P. 2002, ApJ, 579, 386

\bibitem[Rajaniemi \& M\"{a}h\"{o}nen 2002]{rm02} Rajaniemi, H.J. \&
M\"{a}h\"{o}nen, P. 2002, ApJ, 566, 202

\bibitem[Ramirez-Ruiz \& Fenimore 2000]{rafe00} Ramirez-Ruiz, E. \&
Fenimore, E.E. 2000, ApJ, 539, 712

\bibitem[\v{R}\'{\i}pa et al. 2009]{ri09} \v{R}\'{\i}pa, J., 
M\'esz\'aros, A., Wigger, C., Huja, D., Hudec, R. \& Hajdas, W. 
 2009, A\&A, 498, 399

\bibitem[Trumpler \& Weaver 1953]{TW53} Trumpler, R.J. \& Weaver,
H.F. 1953, Statistical Astronomy, University of California Press,
Berkeley

\end{thebibliography}
\end{document}